\pdfoutput=1
\documentclass[a4paper,11pt]{article}
\usepackage{pos}
\usepackage{subcaption}
\usepackage{siunitx}

\begin{document}
\title{ 
Implementing noise reduction techniques into the OpenQ*D package}
%% \ShortTitle{...}
\author*[f]{Lucius Bushnaq}
\author[b]{Isabel Campos}
\author[a]{Marco Catillo}
\author[c]{Alessandro Cotellucci}
\author[d,e,h]{Madeleine Dale}
%\author[f]{Patrick Fritzsch}
\author[c,g]{Jens Lücke}
\author[a]{Marina Krstić Marinković}
\author[c,g]{Agostino Patella}
\author[f]{Mike Peardon}
\author[d,e]{Nazario Tantalo}
\affiliation[a]{Institut für Theoretische Physik, ETH Zürich,\\ Wolfgang-Pauli-Str. 27, 8093 Zürich, Switzerland}
\affiliation[b]{Instituto de Física de Cantabria \& IFCA-CSIC,\\ Avda. de Los Castros s/n, 39005 Santander, Spain}
\affiliation[c]{Humboldt Universität zu Berlin, Institut für Physik \& IRIS Adlershof,\\ Zum Großen Windkanal 6, 12489
Berlin, Germany}
\affiliation[d]{Università di Roma Tor Vergata, Dip. di Fisica,\\ Via della Ricerca Scientifica 1, 00133 Rome, Italy}
\affiliation[e]{INFN, Sezione di Tor Vergata,\\ Via della Ricerca Scientifica 1, 00133 Rome, Italy}
\affiliation[f]{School of Mathematics, Trinity College Dublin,\\ College Green, Dublin 2, Ireland}
\affiliation[g]{DESY,\\ Platanenallee 6, D-15738 Zeuthen, Germany}
\affiliation[h]{Department of Physics, University of Cyprus,  1 Panepistimiou Street, 2109 Aglantzia, Nicosia, Cyprus}
\emailAdd{bushnaql@tcd.ie}
\emailAdd{isabel@campos-it.es}
\emailAdd{Marco.Catillo@physik.uni-muenchen.de}
\emailAdd{alessandro.cotellucci@physik.hu-berlin.de}
\emailAdd{m.dale@stimulate-ejd.eu}
\emailAdd{fritzsch@uni-muenster.de}
\emailAdd{jens.luecke@hu-berlin.de}
\emailAdd{marinama@phys.ethz.ch}
\emailAdd{agostino.patella@physik.hu-berlin.de}
\emailAdd{mjp@maths.tcd.ie}
\emailAdd{nazario.tantalo@roma2.infn.it}
\abstract{We present the results of testing a new technique for stochastic noise reduction in the calculation of propagators by implementing it in OpenQ*D for two ensembles with O(a) improved Wilson fermion action, with periodic boundary conditions and pion masses of $437\,\si{\mega\eV}$ and $331\,\si{\mega\eV}$, for the connected vector and pseudoscalar correlators. We find that the technique yields no speedup compared to traditional methods, owning to the failure of its underlying assumption that the spectra of the spatial Laplacian and Dirac operators are sufficiently similar for the technique's purposes.}

\FullConference{The 38th International Symposium on Lattice Field Theory, 2021 Zoom/Gather@Massachusetts Institute of Technology, LATTICE2021 26th-30th July}
\maketitle
\section{Introduction}
The distillation operator \citep{Peardon_2009} has seen great success as a smearing tool. This might lead one to suspect that the low-lying spectrum of the spatial Laplacian has substantial overlap with the low-lying spectrum of the Dirac operator. Since the spatial Laplacian is considerably smaller, its spectrum is easier to calculate than that of the full Dirac operator. Thus, if the overlap between the two operator’s spectra were good enough, it could make sense to substitute the Distillation spectrum for the Dirac spectrum in many applications.

A potential practical use case and test of this notion would be distillation low-mode averaging. In conventional low-mode averaging, one attempts to reduce the stochastic noise on numeric calculations of propagators by calculating the lowest modes of the Dirac operator explicitly, which are assumed to give the dominant contribution to the observable. The propagator is then broken up into a contribution projected into the space of low-lying modes, which is now available directly, and a space of high-lying modes, which is calculated stochastically \citep{DeGrand_2004} \citep{Giusti_2004}. 

Here, the idea is to do the same, but using the distillation operator instead. The distillation operator is formed out of the eigenmodes of the spatial Laplacian
\begin{equation}
\begin{aligned}
&\bigtriangledown^2_{m s, n s^\prime}(t)=-6\delta_{s s^\prime} \delta_{m n}+\delta_{s s^\prime} \sum^3_{j=1} \left(U_j(m,t)\delta_{m+\hat{j}, n}+U^\dag_j(m-\hat{j},t)\delta_{m-\hat{j}, n}\right)\\
&\Box(t)=V(t)V^\dag(t)=\sum^N_{k=1} v^{(k)}(t)v^{(k)\dag}(t)\,,
\end{aligned}
\end{equation}

where $v^{(k)}(t)$ are the eigenvectors, $N$ is the number of eigenvectors calculated, and $V(t)$ is a matrix with the eigenvectors as column entries. Since the operator does not affect the spin space, the eigenvectors come in degenerate sets of four. We chose the spin-orthogonal basis for these degenerate states.

To implement this method for the calculation of the connected vector correlator, we begin by splitting the Hilbert space into the low-lying space of the Laplacian modes, and the "rest-space" at both source and sink by inserting $1-\Box(t)+\Box(t)$
\begin{equation}
\begin{aligned}
\langle \overline{\psi}^A(t) \gamma_\mu& \psi^B(t) \overline{\psi}^B(t^\prime) \gamma_\nu \psi^A(t^\prime)\rangle_F\\
= & \langle \overline{\psi}^A(t) \gamma_\mu \left(1-\Box(t)+\Box(t)\right) \psi^B(t)\overline{\psi}^B(t^\prime) \gamma_\nu\left(1-\Box(t^\prime)+\Box(t^\prime)\right) \psi^A(t^\prime)\rangle_F\\
= &\langle \overline{\psi}^A(t) \gamma_\mu \left(1-\Box(t)\right) \psi^B(t)\overline{\psi}^B(t^\prime) \gamma_\nu\left(1-\Box(t^\prime)\right) \psi^A(t^\prime)\rangle_F\\
&+2\langle \overline{\psi}^A(t) \gamma_\mu \Box(t) \psi^B(t)\overline{\psi}^B(t^\prime)\gamma_\nu\left(1-\Box(t^\prime)\right) \psi^A(t^\prime)\rangle_F\\
&+\langle \overline{\psi}^A(t) \gamma_\mu \Box(t) \psi^B(t)\overline{\psi}^B(t^\prime) \gamma_\nu\Box(t^\prime) \psi^A(t^\prime)\rangle_F\\
=:&C_{rest}(t)+2 C_{mixed}(t)+C_{dist}(t)\,.\\ 
\end{aligned}
\end{equation}
We have used time reversal symmetry to combine the two mixed contributions into one. If we were working with conventional low-mode averaging, and $\Box(t)$ consisted of eigenmodes of the Dirac operator, $C_{mixed}(t)$ would vanish since $\Box$ and the propagator would commute. But here, we need to calculate it on top of the other two contributions. 

Now we perform a one-end insertion to introduce the stochastic sources
\begin{equation}
\begin{aligned}
&C_{dist}(t)=\textrm{tr}[\gamma_5\gamma_\mu V^\dag(t) D^{-1}(t,t^\prime)V(t^\prime) \gamma_\nu\gamma_5 V^\dag(t^\prime){D^{-1}}^\dag(t^\prime,t)V(t)]\\
&C_{rest}(t)=\frac{1}{N_r}\textrm{tr}[\gamma_5\gamma_\mu \left(1-\Box(t)\right) D^{-1}(t,t^\prime) {\eta}^{(r)}(t^\prime){\eta^\dag}^{(r)}(t^\prime)\gamma_\nu\gamma_5\left(1-\Box(t^\prime)\right) {D^{-1}}^\dag(t^\prime,t)]\\
&C_{mixed}(t)=\frac{1}{N_r}\textrm{tr}[\gamma_5\gamma_\mu \Box(t) D^{-1}(t,t^\prime){\eta}^{(r)}(t^\prime) {\eta^\dag}^{(r)}(t^\prime)\gamma_\nu\gamma_5 \left(1-\Box(t^\prime)\right) {D^{-1}}^\dag(t^\prime,t)]\,.
\end{aligned}
\end{equation}
Thus, to calculate $C_{dist}(t)$ we need the vectors $\phi^{(k)}:= D^{-1}(t,t^\prime)v^{(k)}(t^\prime)$ which then form the perambulator \citep{Peardon_2009}. Thus, we need to solve the Dirac equation with the eigenvectors of the spatial Laplacian at the desired source time as sources.

To calculate $C_{rest}(t)$ and $C_{mixed}(t)$, we need
\begin{equation}
\begin{aligned}
&\phi^{(r)}:= D^{-1}(t,t^\prime){\eta}^{(r)}(t^\prime)\\
&\overline{\phi}^{(r)} := D^{-1}(t,t^\prime)\left(1-\Box(t^\prime)\right)){\eta}^{(r)}(t^\prime)\,,
\end{aligned}
\end{equation}
making for $N+2R$ needed solutions to the Dirac equation in total. Any two-point correlation function can then be formed through the contraction of these solutions by calculating the spin structure explicitly, as detailed in \citep{2008} adjusted for the distillation-low-mode-averaging method.

In order for the technique to be worth the cost of the increased number of solutions needed per stochastic source, as well as the cost of calculating the distillation modes themselves, $C_{mixed}(t)$ and $C_{rest}(t)$ need to be small relative to $C_{dist}(t)$.
\section{Implementation and Simulation setup}
The technique was implemented for the connected vector correlator and pseudoscalars in OpenQ*D, using the PRIMME eigensolver \citep{DBLP:journals/corr/WuRS16} to obtain the distillation eigenmodes.

The spatial covariant laplacian is passed to PRIMME as a black-box operator, parallelized with MPI. The solver then returns the specified number of low-lying eigenmodes. This is repeated for the spatial Laplacian on every timeslice of the lattice, for every configuration. No preconditioning for the operator is employed.

We tested the performance of the method on two CLS  data sets with two flavours of dynamical O(a) improved Wilson fermions. The computation used periodic boundary conditions and stochastic wall sources. The determinations of the lattice spacing as well as the other configuration parameters were taken from Refs. \citep{Fritzsch_2012}\,\citep{Morte_2017}.

\begin{table}[ht]
\centering
\begin{tabular}{c|c|c|c|c|c|c}
  Config&$V$&$\beta$&$\kappa$&$m_{\pi}L$&$a\,[\si{\femto\metre}]$&$m_{\pi}\,[\si{\mega\eV}]$\\  
 \hline
 E5 & $64x32^3$ & $5.30$ & $0.13625$& $4.7$&$0.0658(7)(7)$&$437$\\ 
 A5 & $64x32^3$ & $5.20$ & $0.13594$& $4.0$&$0.0755(9)(7)$&$331$
\end{tabular}
\caption{Ensemble parameters}
\label{table:configs}
\end{table}
\normalsize
\section{Results}
The error analysis for all results was performed with jackknifing. Correlations between the errors at different time slices were not corrected for.
\begin{figure}
\centering
\begin{subfigure}{.5\textwidth}
    \centering
    \includegraphics[scale=0.4]{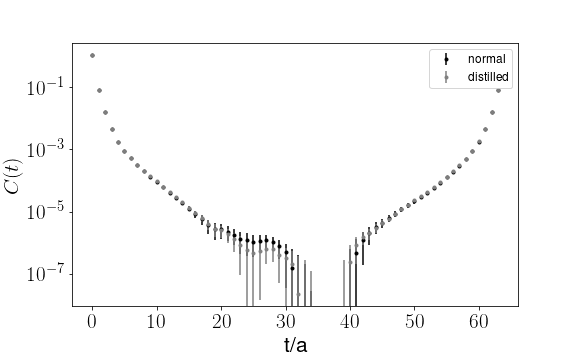}
    \caption{E5}
    \label{fig:corrE5}
\end{subfigure}%
\begin{subfigure}{.5\textwidth}
    \centering
    \includegraphics[scale=0.4]{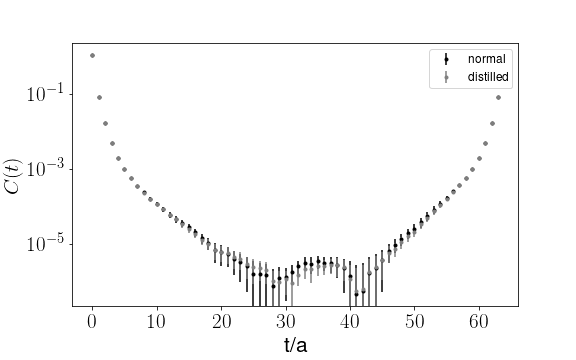}
    \caption{A5}
    \label{fig:corrA5}
\end{subfigure}
\caption{Vector correlator, 160x4 eigenvectors, 24 wall sources}
\label{fig:corr}
\end{figure}
Figures \ref{fig:corrE5} and \ref{fig:corrA5} compare the vector correlator calculated with distillation-low-mode-averaging and without, on the E5 and A5 ensembles respectively, with six independent color and even-odd diluted wall sources and $640$ Laplacian modes. The results are in good agreement inside the error bars.
\begin{figure}
        \centering
        \includegraphics[scale=0.55]{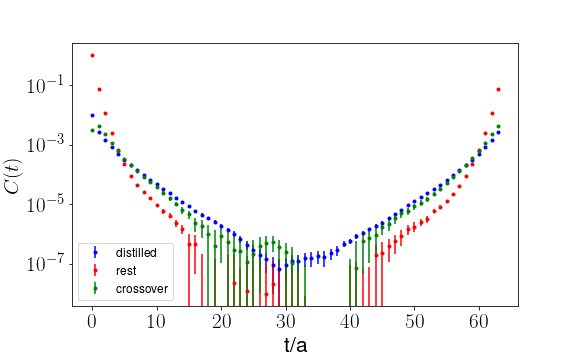}
        \caption{E5 vector correlator by parts, 160x4 eigenvectors, 24 sources}
        \label{fig:E5Vparts}
\end{figure}
Figure \ref{fig:E5Vparts} shows the three individual contributions to the total distilled result in the E5 data set. As one would hope, the $C_{rest}$ contribution quickly falls off in importance relative to $C_{dist}$. However, $C_{mixed}$ does not fall off to the same extent, and stays close to $C_{dist}$ in size even in the long distance part of the correlator.

This is a fundamental problem for the technique. If $C_{mixed}$ does not fall off quickly, the method is not efficient in terms of the computing costs invested. This would seem to indicate that our guess was incorrect, and the spectra of the distillation and Dirac operators are not similar enough to make this idea less numerically expensive after all.
\begin{figure}
\centering
\begin{subfigure}{.5\textwidth}
    \centering
    \includegraphics[scale=0.4]{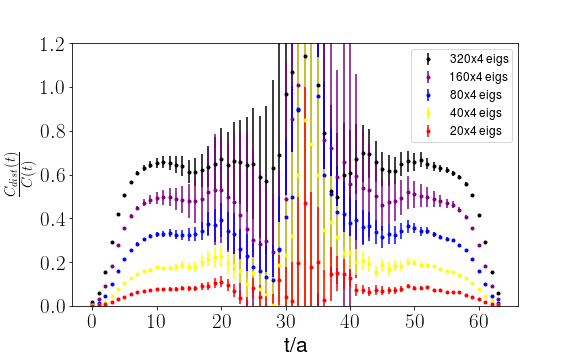}
    \caption{E5}
    \label{fig:percentE5}
\end{subfigure}%
\begin{subfigure}{.5\textwidth}
    \centering
    \includegraphics[scale=0.4]{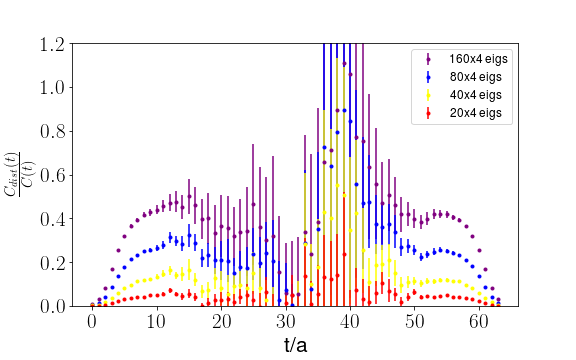}
    \caption{A5}
    \label{fig:percentA5}
\end{subfigure}
\caption{Vector correlator, $\frac{C_{dist}}{C}$ by number of distillation modes, 24 sources}
\label{fig:percent}
\end{figure}
To show this clearly, plots \ref{fig:percentE5} and \ref{fig:percentA5} graph $\frac{C_{dist}}{C}$, the contribution of the distillation sub-space to the whole correlator on the two data sets for different numbers of distillation modes. At low numbers of modes, $\frac{C_{dist}}{C}$ scales linearly with the size of the subspace. But this scaling quickly falls off, and even a computational investment of $1280$ distillation modes falls short of the $0.8$ mark on both ensembles.

\begin{figure}
\centering
\includegraphics[scale=0.55]{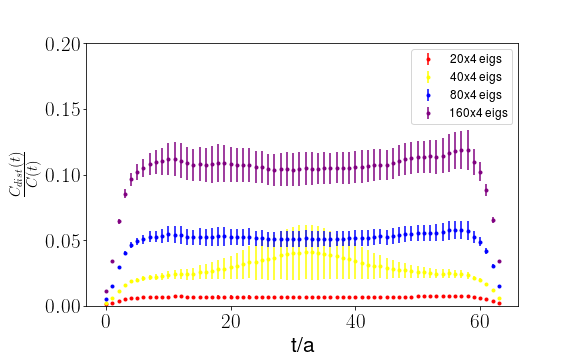}
\caption{A5 Pseudoscalar correlator, $\frac{C_{dist}}{C}$ by number of distillation modes, 24 sources}
\label{fig:percentA5CP}
\end{figure}

The A5 ensemble also exhibits worse scaling overall than E5. We speculate that this is due to the larger size of the physical volume increasing the number of distillation modes below any given threshold \citep{Morningstar_2011}.
\begin{figure}
\centering
\begin{subfigure}{.5\textwidth}
    \centering
    \includegraphics[scale=0.4]{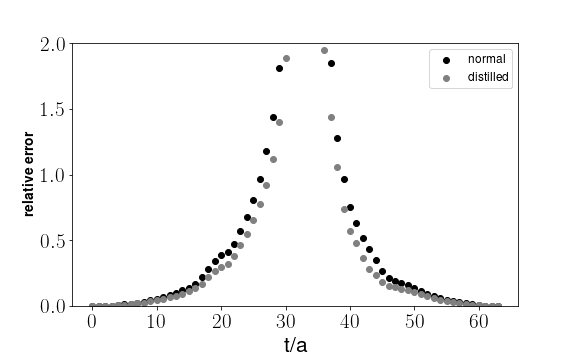}
    \caption{E5}
    \label{fig:precisionE5}
\end{subfigure}%
\begin{subfigure}{.5\textwidth}
    \centering
    \includegraphics[scale=0.4]{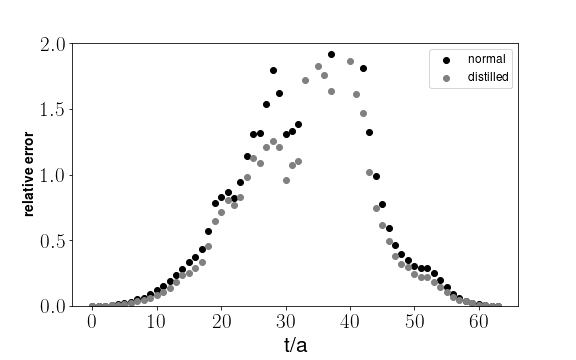}
    \caption{A5}
    \label{fig:precisionA5}
\end{subfigure}
\label{fig:precision}
\caption{Vector correlator, $\frac{\sigma_{C}}{C}$, 160x4 eigenmodes, 24 sources}
\end{figure}
Figures \ref{fig:precisionE5} and \ref{fig:precisionA5} plot the relative errors $\frac{\sigma_C}{C}$ of the total correlators with and without the distillation technique applied, for 160 degenerate distillation modes. As might be expected from the analysis above, there is a factor of about $\frac{1}{\sqrt{2}}$ between the two, which is consistent with reducing the stochastic noise on half of the total correlator to approximately zero, since
\begin{equation}
\begin{aligned}
&\sigma_C=\sqrt{\sigma^2_{C_{dist}}+\sigma^2_{C-C_{dist}}}\,.
\end{aligned}
\end{equation}
This means that if the contribution of $\sigma_{C-C_{dist}}$ to the total stochastic error budget simply scaled linearly with the size of $C-C_{dist}$, one would expect a $\frac{1}{\sqrt{2}}$ factor for $\frac{C_{dist}}{C}=0.5$, which is what we observe. Similar results are observed for the pseudoscalar correlator, except that the scaling of $\frac{C_{dist}}{C}$ with the number of modes is even more unfavourable. This is pictured in figure \ref{fig:percentA5CP}
\subsection{Computational costs}
To compare computational costs, table \ref{table:costs} shows the time needed to calculate the distillation eigenmodes with Primme, and $C_{dist}$ with OpenQ*D, on a single configuration, for different numbers of eigenmodes.
\begin{table}[ht]
\centering
\scriptsize
\begin{tabular}{c|c|c|c|c}
 $N_{dist}$ &$C_{dist}$ Dirac solver&Lap. solver&Lap. solver Ortho.&Lap. solver MatVec\\  
 \hline
 $20x4$ & $1.18\mathrm{e}{+02}\,\text{s}$ & $46.3\,\text{s}$ & $3.5\,\text{s}$& $31.1\,\text{s}$\\ 
 $40x4$ & $2.53\mathrm{e}{+02}\,\text{s}$ & $1.04\mathrm{e}{+02}\,\text{s}$ & $10.8\,\text{s}$& $68.7\,\text{s}$\\  
 $80x4$ & $4.84\mathrm{e}{+02}\,\text{s}$ & $2.43\mathrm{e}{+02}\,\text{s}$ & $41.8\,\text{s}$& $1.52\mathrm{e}{+02}\,\text{s}$\\  
 $160x4$ & $1.12\mathrm{e}{+03}\,\text{s}$ & $7.77\mathrm{e}{+02}\,\text{s}$ & $2.51\mathrm{e}{+02}\,\text{s}$& $3.46\mathrm{e}{+02}\,\text{s}$\\  
 $320x4$ & $2.82\mathrm{e}{+03}\,\text{s}$ & $2.14\mathrm{e}{+03}\,\text{s}$ & $1.14\mathrm{e}{+03}\,\text{s}$& $7.40\mathrm{e}{+02}\,\text{s}$\\  
\end{tabular}
\caption{Computational costs}
\label{table:costs}
\end{table}
At low numbers of distillation modes, the computational cost is dominated by the Dirac solver. However, as the number of modes is increased, the time needed by the Primme Laplacian solver for orthogonalisation increases quadratically.

This represents another fundamental obstacle for the technique. Since a very large number of eigenmodes seems needed for $C_{dist}$ to dominate $C_{mixed}$ in the correlator, the time needed to solve for all the eigenmodes of the laplacian will eventually scale quadratically with the number of modes, potentially destroying the supposed advantage over normal low-mode-averaging that individual modes take less time to calculate.

If we were to use the computation time invested in distillation for additional stochasic sources instead, we would expect
\begin{equation}
\begin{aligned}
&t_{\frac{N}{4}, l}+t_{N, D}+2t_{R_1, D}= t_{R_0, D}\,,\\
\end{aligned}
\end{equation}
where $R_1$ is the number of stochastic sources with distillation, $R_0$ the number possible for the same cost without distillation, $t_{N, D}$ the time required to solver the Dirac equation for the given number of distillation modes $N$, $t_{R, D}$ the time required to do the same for the stochastic sources and $t_{\frac{N}{4}, l}$ the time required to run the Laplacian solver for the given number of distillation modes $N$. The factor of $2$ reflects the fact that introducing distillation forces us to solve the Dirac equation for the stochastic sources twice, with different preconditionings.

Assuming roughly equal time needed to solve the equation for any kind of Dirac source, this implies
\begin{equation}
\begin{aligned}
R_0 > 2R_1+N\,,\\
\end{aligned}
\end{equation}

since the relative variance of the correlator scales as
\begin{equation}
\begin{aligned}
&\frac{{\sigma_C}^2}{C} \propto \frac{1}{R}\,,
\end{aligned}
\end{equation}
we would thus need 
\begin{equation}
\begin{aligned}
&1-\frac{C_{dist}}{C} \leq \frac{R_1}{R_0} \leq  \frac{R_1}{2R_1+N} = \frac{1}{\frac{N}{R_1}+2}\,,
\end{aligned}
\end{equation}
even in the limit of assuming the time needed to find the Laplacian modes is negligible, which is clearly not the case.

In our existing tests plotted in figures \ref{fig:percentE5} and \ref{fig:percentA5}, this relation is never fulfilled.
\section{Conclusion}
We conclude that distillation-low-mode-averaging does not seem computationally efficient for the connected vector and pseudoscalar correlators.  Contrary to our hopes, the similarity between the low-lying spectra of the spatial Laplacian and the Dirac operator was not strong enough to allow efficiently substituting the eigenmodes of the former for the later. Since the eigenmodes of the spatial Laplacian become more dense at larger lattice volumes, we would expect the efficiency of the technique to fall off further if the physical lattice volume were increased.

If the underlying idea of finding a more computationally friendly substitute for the Dirac operator in low-mode-averaging were to be pursued further, a possible next step could be testing operators with a non-trivial spin structure.
\section{Acknowledgements}
This project has received funding from the European Union’s Horizon 2020 research and innovation programme under the Marie Skłodowska-Curie grant agreement No 813942. The research of AC, JL and AP is funded by the Deutsche Forschungsgemeinschaft (DFG, German Research Foundation) - Projektnummer 417533893/GRK2575 “Rethinking Quantum Field Theory”. The work was supported by the North-German Supercomputing Alliance (HLRN) with the project bep00085. The work was supported by the Poznan Supercomputing and Networking Center (PSNC) through grant numbers 450 and 466. The authors acknowledge access to Piz Daint at the Swiss National Supercomputing Centre, Switzerland under the ETHZ’s share with the project IDs go22 and go24.  Computations were also performed on the Seagull cluster maintained by the Trinity Centre for High Performance Computing (TCHPC).
\bibliographystyle{JHEP}
\bibliography{Proceedings_Lattice_2021} 
\end{document}